\documentclass[preprint,12pt]{elsarticle}

\usepackage{graphicx}
\usepackage{dcolumn}
\usepackage{amsmath}
\usepackage{amssymb}
\usepackage{bm}
\usepackage{lineno}
\usepackage{geometry}
\usepackage{color}
\usepackage{siunitx}
\usepackage{mathtools}
\usepackage{tabularx}
\usepackage{booktabs}
\usepackage{multirow}
\DeclareSIUnit\angstrom{\text {Å}}

\journal{xxx}

\begin{document}

\begin{frontmatter}



\title{The Performance of Seeded Free-Electron Lasers Through Dispersion Strength Tuning}
\author[add1]{Li Zeng}
\author[add2]{Chao Feng\corref{cor1}}\ead{fengc@sari.ac.cn}
\author[add1]{Xiaofan Wang\corref{cor1}}\ead{wangxf@mail.iasf.ac.cn}
\author[add1]{Huaiqian Yi}
\author[add3]{Weiqing Zhang\corref{cor1}}
\ead{weiqingzhang@dicp.ac.cn}
 
\cortext[cor1]{Corresponding authors.}
\address[add1]{Institute of Advanced Science Facilities, Shenzhen 518107, China}
\address[add2]{Shanghai Advanced Research Institute,Chinese Academy of Sciences, Shanghai 201210, China}
\address[add3]{Dalian Institute of Chemical Physics, Chinese Academy of Sciences, Dalian 116023, China}

\begin{abstract}
Over the last decade, external seeded free electron lasers (FELs) have achieved significant advancements across various disciplines, progressively establishing themselves as indispensable tools in fields ranging from fundamental science to industrial applications. 
The performance of seeded FELs is critically dependent on the quality of the frequency up-conversion process. Optimized conditions for seeded FELs are typically considered as the maximization of the bunching factor. 
This paper discusses alternative perspectives on the optimization criteria for seeded FELs by analyzing the impact of dispersion strength on their overall performance. We investigate the relationship among the required dispersion strength for achieving the maximum bunching factor, maximum pulse energy, and optimal energy stability through theoretical analysis, simulation calculations, and experimental explorations. Additionally, the direct observation of pulse splitting emphasizes the consideration of trade-off between pulse energy and temporal coherence in seeded FELs. These results provide valuable insights and practical guidance for controlling the pulse characteristics of seeded FELs, contributing to the tuning and optimization of FEL facilities.
\end{abstract}


\end{frontmatter}


\section{Introduction}\label{introduction}

The revolutionary changes introduced by Free-electron Lasers (FELs) have significantly enhanced our ability to explore complex materials and biological systems with unprecedented precision and versatility, paving the way for groundbreaking discoveries across multiple fields, including medicine, materials science, and fundamental physics~\cite{Ostrom2015Probing, Gerber2015Three, Rudenko2017femtosecond}. Currently, several FEL facilities have been constructed worldwide~\cite{Ackermann2007Operation, Emma2010First, Ishikawa2012A, Allaria2012Highly, Kang2017Hard, Prat2020A, Chao2022Coherent}, serving as dispensable tools for a wide variety of scientific research. Of particular interest among these studies is the experiment implemented at seeded FELs, which has the major advantage of full coherence, precise control over arrival time, and a uniform longitudinal profile.

The capability to produce stable, fully coherent radiation pulses has garnered sustained attention for external seeded FELs, which employ optical-scale electron beam manipulation with external seed lasers. The typical operation modes of external seeded FELs are high-gain harmonic generation (HGHG)~\cite{Yu1991Generation, Yu2000high, Yu2003first} and echo-enabled harmonic generation (EEHG)~\cite{Stupakov2009Using, Xiang2009Echo, Zhao2012First}. In the HGHG scheme, an external seed laser interacts with the electron beam in the modulator, imprinting a sinusoidal energy modulation onto the longitudinal phase space. After passing through a dispersion chicane, this energy modulation is converted into a density modulation (micro-bunching), which contains frequency components at high harmonics of the seed laser. The desired coherent radiation is then generated and amplified in the subsequent undulator by this micro-bunched electron beam. In the EEHG scheme, two modulation-dispersion sections induce successive transformations of the electron beam longitudinal phase space, accumulating electrons into fine energy bands and leading to multiple density spikes within each seed wavelength. Other proposals, such as phase-merging enhanced harmonic generation (PEHG)~\cite{Deng2013Using, Feng2014Phase} and direct-amplification enabled harmonic generation (DEHG)~\cite{Wang2022High}, have not been experimentally validated due to the complexity of their required machine architectures.

Building on the above introduction, it is evident that the frequency up-conversion process, which dominates the generation of bunching factor, plays a crucial role in determining the performance of seeded FELs. Once temporal and spatial synchronization between the seed laser and the electron beam is established~\cite{Liu2013Precise, Lechner2014Measurement, Cinquegrana2021The}, adjusting the magnet currents in the dispersion chicane is generally required to optimize the electron beam bunching factor. Indeed, one generally identifies the optimized conditions for seeded FEL emission as those that maximize the bunching factor~\cite{Finetti2017Pulse, Giannessi2022FERMI, Feng2016A}. However, to ensure the stable operation of FELs and the successful execution of user experiments, it is more critical to meet stringent specifications, particularly in terms of pulse energy and its stability as well as the spatial, temporal, and spectral characteristics of the FEL pulses~\cite{Bettoni2021Overview, Deng2014Simulation, Justus2008Investigation}. Therefore, optimizing the parameters of the harmonic conversion stage to maximize pulse energy or achieve optimal energy stability, rather than merely maximizing the bunching factor as suggested in Ref.~\cite{Yang2022Optimization, Wang2016Study}, appears to be a more significant issue for consideration. Meanwhile, the evolution of other radiation pulse characteristics, such as power profile, spectrum and time-bandwidth product (TBP), under these newly optimized parameters requires further investigation.

This paper explores an alternative optimization criterion for seeded FELs by analyzing the impact of dispersion strength on the performance of FEL pulses. The study investigates optimal conditions and FEL performance with respect to the maximum bunching factor, pulse energy, and energy stability through theoretical calculations, numerical simulations, and experimental observations. Furthermore, the direct experimental observations reveal the phenomena of pulse splitting and corresponding spectral splitting~\cite{Finetti2017Pulse, Gauthier2015Spectrotemporal, labat2009pulse, Mirian2020Spectrotemporal}, which are subsequently explained through theoretical analysis. The results provide new insights into the efficient control of FEL pulse characteristics and contribute to the tuning and optimization of seeded FELs.

\section{Theoretical Analysis}\label{sec2:theory}

In seeded FELs, the electron beam is spatially bunched through longitudinal phase space manipulations. This bunched beam contains frequency components at high harmonics of the seed laser, which significantly enhances the radiation at the desired wavelengths. With this kind of electron beam, the lasing process before the saturation can be broadly categorized into two stages.

In the first two gain lengths of the radiator, the FEL operates in the coherent harmonic generation (CHG) regime. Within this regime, the harmonic field increases linearly with the distance $z$ traveled through the radiator, while the power grows quadratically with $z$. The radiation power at the end of this regime can be calculated as~\cite{Yu2002Theory, Wu2004X}
\begin{equation}
P_{coh} = \frac{I_0^2Z_0}{8}\left(\frac{K[JJ]}{\gamma}\right)^2\frac{1}{4\pi\sigma_x^2}\left|I_b\right|^2,
\label{eq1}
\end{equation}
where $I_0$ and $\sigma_x$ represent the electron beam current and the rms beam size, respectively. Here, $Z_0=377\Omega$ denotes the vacuum impedance, while $K$ and $[JJ]$ correspond to the undulator parameter and the Bessel factor, respectively. $I_b$ is defined as the integral of the bunching factor over two gain lengths:
\begin{equation}
I_b=\int_0^{2L_G}b_n(z)dz.
\label{eq2}
\end{equation}
In case of idealized model, this yields  $I_b\approx2L_{G}b_n$.

After two gain lengths, the lasing process enters the exponential growth region, where the radiation power can be estimated as:
\begin{equation}
P(z) = CP_{coh}e^{z/L_{G}},
\label{eq3}
\end{equation}
where $C$ is the coupling coefficient, which is approximately $1/3$ for a large electron beam size estimation~\cite{Wu2004X}. The equation indicates that the coherent harmonic radiation in the first two gain lengths, serving as the start-up power, would coupled into the guided mode with a coefficient of $C$. Subsequently, this power is exponentially amplified further in the undulator. 

The saturation power can be expressed as $P_{sat}=CP_{coh}e^{L_{sat}/L_{G}}$, where $L_{sat}$ denotes the saturation length. Consequently, for an ideal electron beam, the pulse energy at saturation is proportional to the integral of the square of bunching factor over the entire beam, represented as $PE\propto\int b_{n}^2ds$. To achieve maximum pulse energy, it is essential to optimize the bunching factor to maximize $\int b_{n}^2ds$, rather than simply maximizing $b_{n}$.

For the HGHG, the amplitude of bunching factor at the $n$th harmonic of the seed laser, evaluated at the longitudinal coordinate $\zeta$, can be written as
\begin{equation}
b_n(\zeta)=J_n(nA(\zeta)B)e^{-\frac{n^2B^2}{2}},
\label{eq4}
\end{equation}
where $J_n$ is the Bessel function of the first kind of order $n$, $A=\Delta\gamma/\gamma$ represents the dimensionless energy modulation amplitudes induced by the seed laser, and $B=R_{56}k_{s}\sigma_{\gamma}/\gamma$ denotes the dimensionless strength of the dispersion section. 

\begin{figure}[!htb]
  \centerline{
  \includegraphics[width=0.93\textwidth]{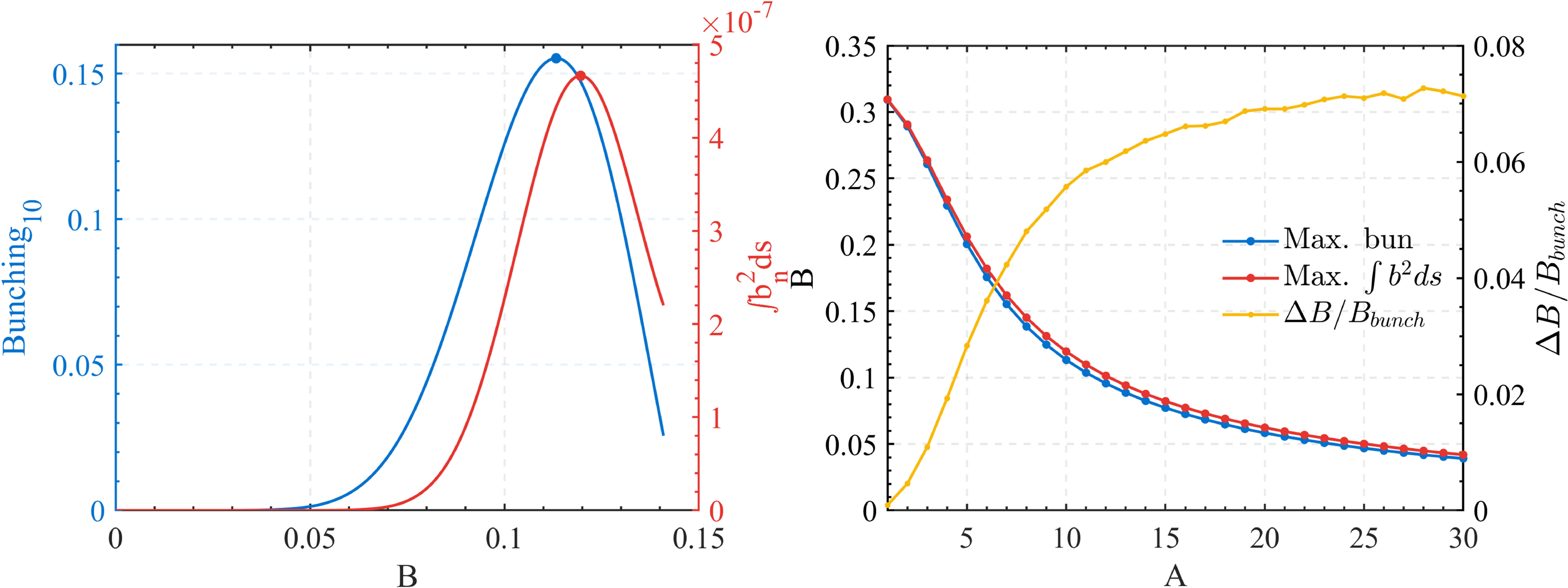}}
   \caption{Variation of maximum bunching factor (blue) and the integral of $b_{n}^2$ (red) as a function of $B$ under the conditions of $A=10$ and $n=10$ for the HGHG (left panel). The dimensionless dispersion strength required to achieve the maximum bunching factor (blue) and the maximum value of $\int b_{n}^2ds$ (red) along with their differences (yellow), for varying values of $A$ (right panel).}
   \label{fig1:theory1}
\end{figure}

\begin{figure}[!htb]
  \centerline{
  \includegraphics[width=0.48\textwidth]{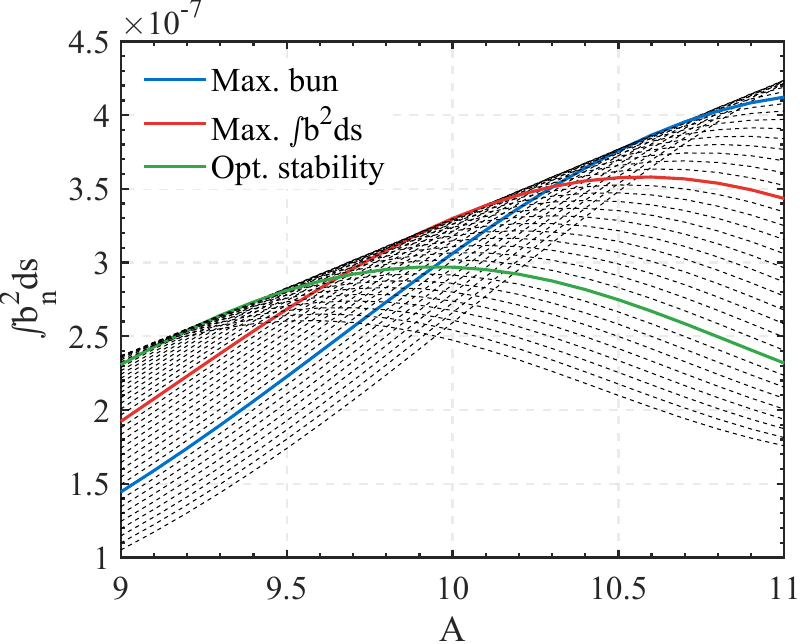}}
   \caption{The integral of $b_{n}^2$ for various dispersion strengths, with a $\pm$10\% deviation in $A$, for the HGHG. The blue, red and green lines represent the conditions of maximum bunching, maximum $\int b_{n}^2ds$ and optimal pulse energy stability, respectively.}
   \label{fig2:theory2}
\end{figure}

Assuming a seed laser with a pulse length (FWHM) of \SI{100}{fs} that induces an energy modulation of $A=10$, Figure~\ref{fig1:theory1} illustrates the evolution of the maximum harmonic bunching factor at the 10$^{th}$ harmonic (blue) and the integral of $b_{n}^{2}$ as the dispersion strength $B$ is varied. The required dispersion strengths for achieving the maximum bunching factor ($B_{bunching}$) and the maximum integral of $b_{n}^{2}$ ($B_{integral}$), along with their differences, are also depicted for various modulation amplitudes $A$. The figure indicates that $B_{bunching}$ is consistently smaller than $B_{integral}$, suggesting a slight over-tuning of the dispersion chicane would result in a higher FEL pulse energy. Figure~\ref{fig2:theory2} demonstrates the integral of $b_{n}^{2}$ across various dispersion strength with a $\pm$10\% deviation in energy modulation amplitude. The projections of these curves along the y-axis reflect the fluctuations of the pulse energy. For optimal pulse energy stability, the required dispersion strength ($B_{stability}$) should be slightly greater than $B_{integral}$.

For the EEHG, the bunching factor is expressed as
\begin{equation}
b_{n,m}(\zeta)=e^{-\frac{\xi_{E}^2}{2}}J_n(-\xi_{E}A_{1}(\zeta))J_m(-a_{E}A_2(\zeta)B2),
\label{eq5}
\end{equation}
where $\xi_{E}=nB_1+a_{E}B_2$ represents the EEHG scaling parameter, and the harmonic number $a_E=n+mK$ with $n$ and $m$ as integers and $K=\lambda_{1}/\lambda_{2}$. The terms $A_{1,2}$ and $B_{1,2}$ denote the laser modulation amplitudes and dispersion strengths, respectively.

\begin{figure}[!htb]
  \centerline{
  \includegraphics[width=0.93\textwidth]{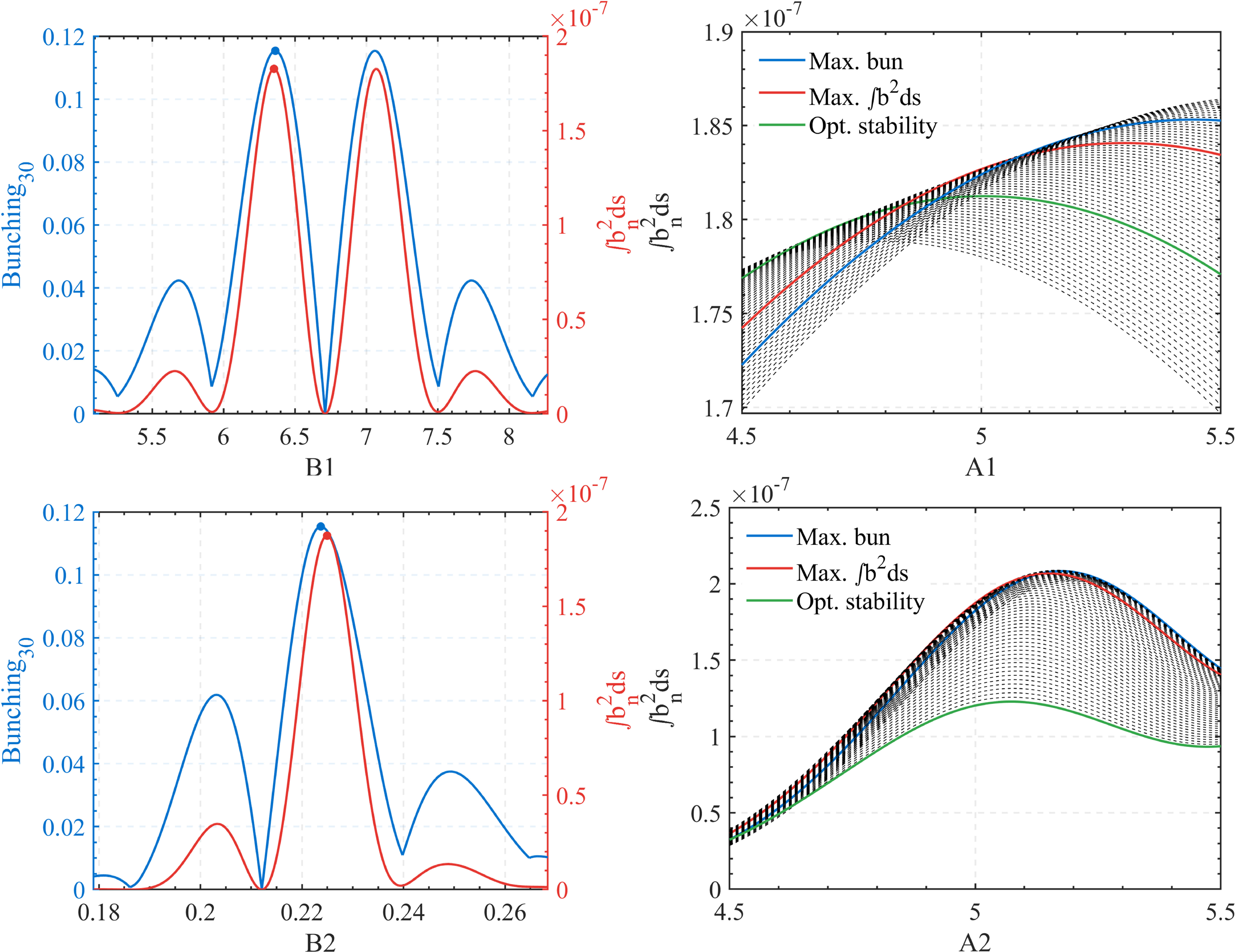}}
   \caption{The influence of $B1$ (top) and $B2$ (bottom) on the maximum bunching factor, the integral of $b_{n}^2$ and pulse energy stability for the EEHG.}
   \label{fig3:theory3}
\end{figure}

Two identical seed lasers, each with a pulse duration of \SI{100}{fs}, are employed to induce energy modulations of $A_1=A_2=5$. The harmonic number is set to be 30, with $n$ and $m$ assigned as -1 and 31, respectively. To enhance clarity in the analysis, the effects of the two modulation-dispersion sections are considered separately. Figure.~\ref{fig3:theory3} depicts the effects of the first (top) and second (bottom) dispersion strengths on the maximum bunching factor, the integral of $b_n^{2}$, and pulse energy stability. In the first modulation-dispersion section, the dispersion strengths follow the relationship $B_{bunching} > B_{integral} > B_{stability}$, which contrasts sharply with observations in HGHG. In the second modulation-dispersion section, $B_{integral}$ is marginally greater than $B_{bunching}$. As noted in Ref.~\cite{Hemsing2017Sensitivity}, a $\pm$10\% deviation in $A_2$ significantly impacts the formation of the bunching factor at $m=31$; consequently, the strength of $B_{stability}$ will not be considered. 

The analysis suggests that, instead of exclusively pursuing maximum bunching factors, a slight over-tuning or under-tuning of the dispersion section in seeded FELs can increase pulse energy and improve energy stability. Furthermore, such tuning can smooth the transverse profile of FEL radiation considering variations in energy modulation across the electron beam's transverse section. This smoothing mechanism may lead to a larger Rayleigh region, thereby reducing the diffraction loss in the subsequent transport line. To improve conciseness while maintaining generality, the following simulations and experimental studies will primarily concentrate on the impact of dispersion strength on the performance of HGHG FELs.

\section{Simulation Results}\label{sec3:simulation}

To comprehensively investigate the influence of dispersion strength, simulations were conducted based on the parameters specified in Table~\ref{tab1:par}. A flat-top electron beam with a full bunch length of approximately \SI{230}{fs} is adopted. The electron beam energy is set to \SI{2.5}{GeV} with a relative energy spread of \SI{8e-5}{}. The normalized emittance is \SI{0.5}{mm\cdot mrad}, and the peak current is \SI{800}{A}. A \SI{270}{nm} seed laser with a pulse duration of \SI{100}{fs} (FWHM) imprints a sinusoidal energy modulation onto the electron beam as it passes through a 2-meter-long modulator with a period of \SI{0.09}{m}. The peak power of the seed laser is approximately \SI{410}{MW}, resulting in a dimensionless energy modulation of about 10.  Following the dispersion chicane, this energy modulation is converted into a density modulation that contains the 10th harmonic component. The electron beam then enters the radiator section, which consists of six 4-meter-long undulators with a period of \SI{0.05}{m}. After adjusting the undulator parameters to meet the resonant condition, coherent \SI{27}{nm} radiation pulses are generated and amplified in the radiator.

\begin{table}[!htb]
 \caption{Simulation~parameters.\label{tab1:par}}
 \newcolumntype{C}{>{\centering\arraybackslash}X}
 \begin{tabularx}{\textwidth}{CCCC}
 \toprule
 \textbf{Section} & \textbf{Parameter} & \textbf{Value} & \textbf{Unit} \\
 \midrule
 \multirow[m]{4}{*}{Electron beam} & Beam energy & 2.5 & GeV\\
                        & Energy spread & $8\times10^{-5}$ & --\\
                        & Emittance & 0.5/0.5 & mm$\cdot$mrad\\
			 	    & Peak current & 800 & A\\
 \midrule
 \multirow[m]{3}{*}{Seed laser} & Wavelength & 270 & nm\\
			 	    & Pulse length & 100 & fs\\
			   	  & Peak Power & $\sim$410 & MW\\
 \midrule
 \multirow[m]{2}{*}{Modulator} & Period length & 0.09 & m\\
			 	    & Total length & $\sim$2 & m\\
 \midrule
 \multirow[m]{3}{*}{Radiator} & Period length & 0.05 & m\\
			 	    & Undulator length & 4 & m\\
			   	  & FEL wavelength & 27 & nm\\
 \bottomrule
 \end{tabularx}
\end{table}

\begin{figure}[!htb]
  \centerline{
  \includegraphics[width=0.6\textwidth]{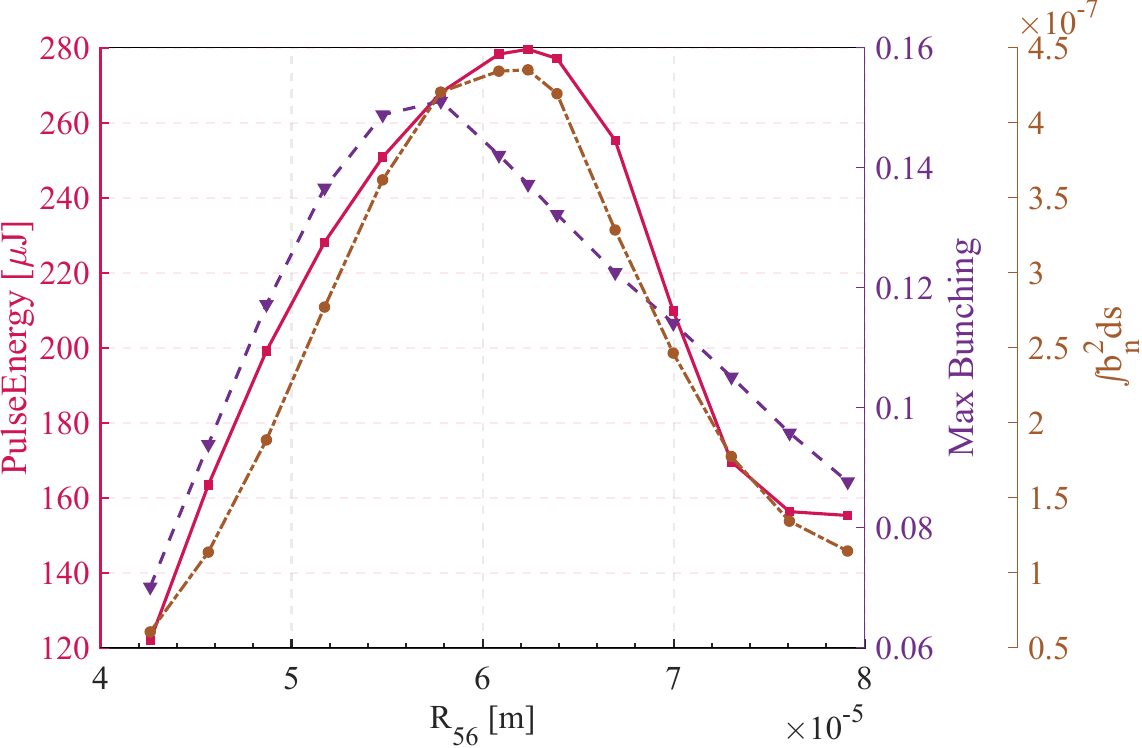}}
   \caption{The pulse energy (red), maximum bunching factor before radiator (purple) and the integral of  $b_{n}^2$ for various dispersion strengths.}
   \label{fig4:sim1}
\end{figure}

The simulations were performed with \textit{GENESIS4}~\cite{Reiche2014Update}. Figure \ref{fig4:sim1} illustrates the simulated pulse energy at saturation (red), the maximum bunching factor before the radiator (purple), and the integral of $b_n^{2}$ as a function of dispersion strength. These characteristics experience an initial growth followed by a subsequent decline as the dispersion strength increases. Notably, the dispersion strength at which maximum pulse energy occurs ($R_{56}^{pe}=\SI{6.24e-5}{m}$) is larger than that corresponding to the maximum bunching factor ($R_{56}^{bun}=\SI{5.78e-5}{m}$). Moreover, the pulse energy exhibits a trend that closely aligns with the integral of $b_n^{2}$ (particularly $R_{56}^{pe}=R_{56}^{int}$), further substantiating the theoretical analysis presented in Section~\ref{sec2:theory}.

\begin{figure}[!htb]
  \centerline{
  \includegraphics[width=\textwidth]{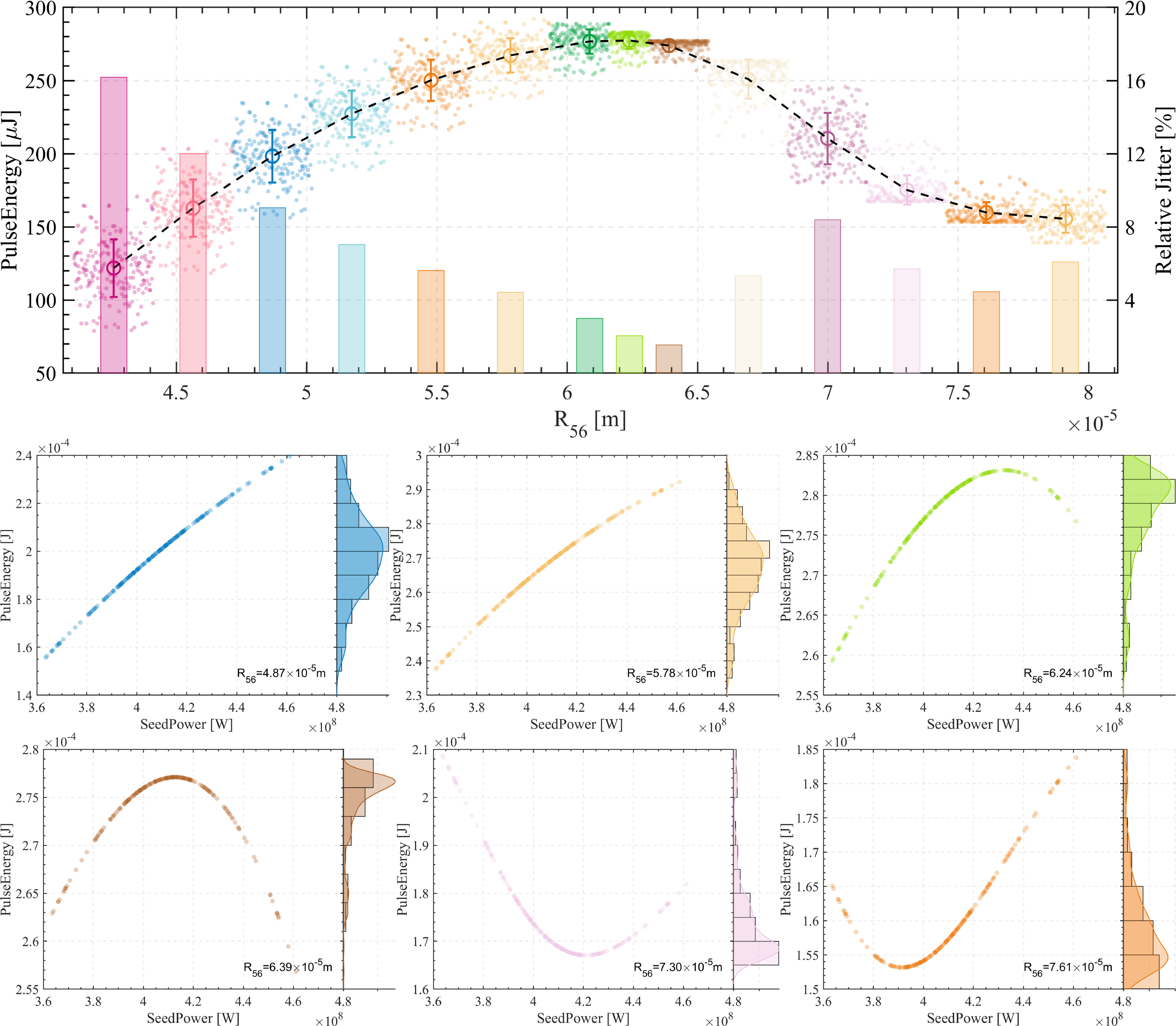}}
   \caption{Simulated pulse energies at different dispersion strengths. The histogram illustrates the relative jitter of pulse energy. The lower panel displays the correlation between seed laser powers and FEL pulse energies at selected dispersion strengths.}
   \label{fig5:sim2}
\end{figure}

Numerous simulations were conducted to investigate the influence of dispersion strength on pulse energy stability. Each dispersion condition was simulated 200 times, incorporating a 5\% RMS jitter in the seed laser power. Figure~\ref{fig5:sim2} provides a comprehensive overview of these simulations. The upper panel illustrates the pulse energy across varying dispersion strengths, while the accompanying histogram displays the relative jitter of pulse energy. To achieve optimal pulse energy stability, a dispersion strength ($R_{56}^{sta}$) of approximately \SI{6.39e-5}{m} is required, which is slightly larger than the condition that yields maximum pulse energy. Combining the results presented in Fig.~\ref{fig4:sim1}, it's evident that $R_{56}^{sta}>R_{56}^{pe}>R_{56}^{bun}$, which is fully consistent with the theoretical findings. 

Furthermore, the relative jitter of pulse energy exhibits an initial decrease, followed by a sharp increase, and then another decline before rising again as the dispersion strength intensifies. This variation is attributed to presence of the Bessel function in Eq.~(\ref{eq4}). Specifically, the first local minimum of the jitter corresponds to the first local maximum of the Bessel function, while the second local minimum is associated with the first zero of the Bessel function. As shown in the lower panel of Fig.~\ref{fig5:sim2}, it is anticipated that the second local minimum of the jitter will occur between \SI{7.30e-5}{m} and \SI{7.61e-5}{m}, which is in close proximity to the theoretically calculation of \SI{7.77e-5}{m}. This discrepancy arises because the energy modulation amplitude introduced by the seed laser is slightly greater than 10.

\begin{figure}[!htb]
  \centerline{
  \includegraphics[width=0.9\textwidth]{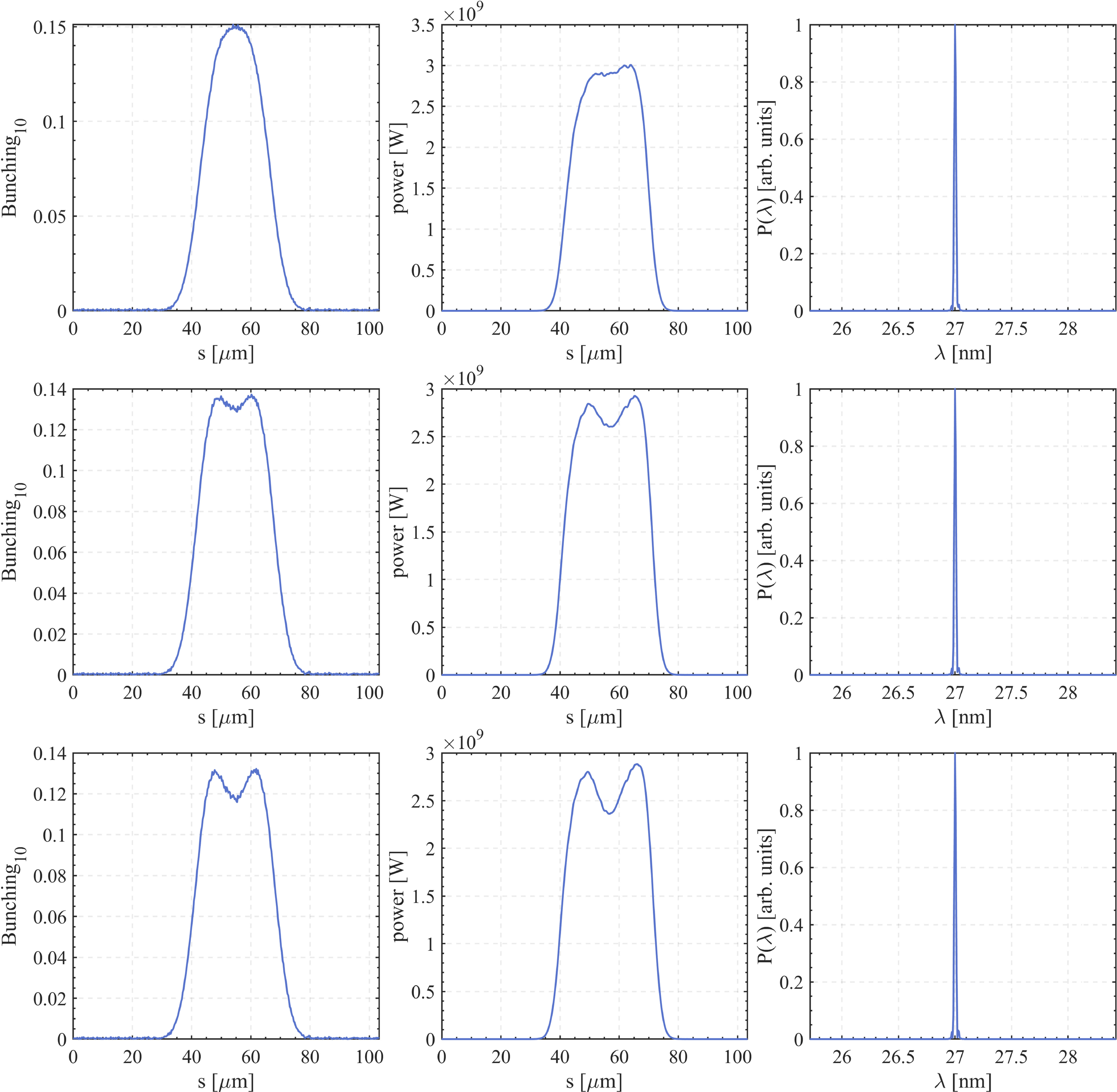}}
   \caption{Simulated 10$^{th}$ harmonic bunching factors, power profiles and spectra under three conditions: maximum bunching (top), maximum pulse energy (middle), and optimal pulse energy stability (bottom).}
   \label{fig6:sim3}
\end{figure}

Figure~\ref{fig6:sim3} demonstrates the simulated 10$^{th}$ harmonic bunching factors, power profiles and spectra at saturation under conditions of maximum bunching (top), maximum pulse energy (middle), and optimal pulse energy stability (bottom). The corresponding dispersion strengths for these conditions are \SI{5.78e-5}{m}, \SI{6.24e-5}{m}, and \SI{6.39e-5}{m}, with the last two showing slight over-tuning and pulse-splitting. The pulse energies for these cases are \SI{268.01}{\micro\joule}, \SI{279.56}{\micro\joule}, and \SI{277.09}{\micro\joule}, while the pulse durations are \SI{91.68}{fs}, \SI{100.12}{fs}, and \SI{102.59}{fs}, respectively. Combining the spectral results, the time-bandwidth products (TBP) are approximately 0.7529, 0.7759, and 0.7785, which correspond to factors of 1.71, 1.76, and 1.77 times the Fourier transform limit, respectively. This suggests that all conditions shows relatively high temporal coherence.

The above simulation results indicate that a slight over-tuning of dispersion strength, without substantially compromising temporal coherence, can enhance the FEL pulse energy and improve the energy stability. This finding provides constructive guidance on the tuning and optimization of seeded FELs.

\section{Experiment Results}\label{experiment}

To further elucidate the influence of dispersion strength on the FEL performance, an experiment was conducted at the Shanghai Soft X-ray FEL Facility (SXFEL)~\cite{Liu2022The}, which can operate in HGHG mode. The electron beam energy is approximately \SI{793}{MeV}, with a relative project energy spread of about \SI{e-3}{}. The electron bunch is compressed to about \SI{1}{ps} (FWHM) using a magnetic chicane, resulting in a peak current exceeding \SI{500}{A}. The normalized emittance after linac is around \SI{1.5}{mm\cdot mrad}. The seed laser is derived from a commercial Titanium:Sapphire laser system, capable of generating output lases of \SI{3}{mJ} at a wavelength of \SI{800}{nm}. Third harmonic generation is subsequently employed to convert this output to about \SI{267}{nm} ($\sim$\SI{170}{fs} (FWHM)). The radiator comprises three variable-gap undulators, each 3 meters long, with a period of \SI{40}{mm}. The installation of an X-band RF deflector in conjunction with a dipole magnet at the end of the radiator section facilitates the acquisition of the electron beam longitudinal phase space images and the reconstruction of FEL power profiles. The FEL spectra are obtained with a dedicated VUV spectrometer, which has a spectral coverage from \SI{40}{nm} to \SI{200}{nm} and a resolution of \SI{0.05}{nm}. The temporal resolution is approximately \SI{6}{fs}, while the energy resolution is around \SI{36}{keV}.

\begin{figure}[!htb]
  \centerline{
  \includegraphics[width=\textwidth]{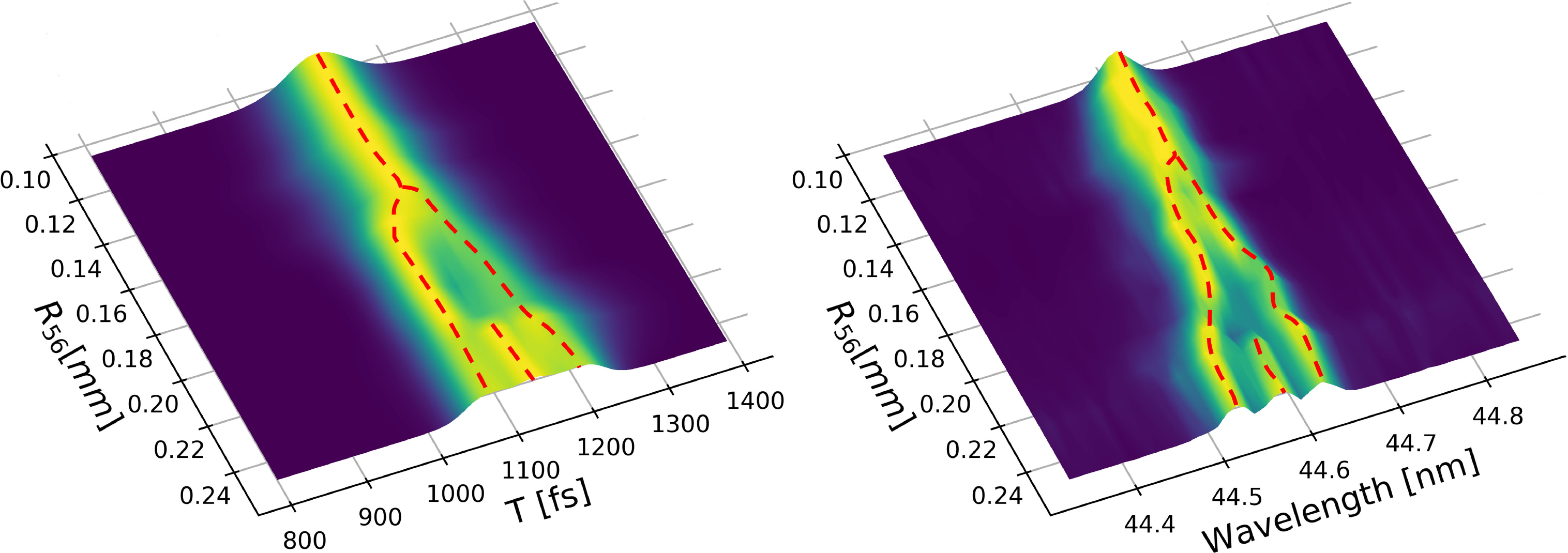}}
   \caption{Experimental measurements of the reconstructed pulse longitudinal profiles (left) and the corresponding spectra measured simultaneously using the spectrometer (right). (The data for each $R_{56}$ have been normalized in amplitude for better visualization.)}
   \label{fig7:exp1}
\end{figure}

During the experiment, the HGHG was optimized for 6$^{th}$ harmonic generation. The current of dispersion chicane magnets was scanned from \SI{20}{A} to \SI{31}{A}, resulting in a corresponding $R_{56}$ range of approximately \SI{0.11}{mm} to \SI{0.25}{mm}. The longitudinal phase space and corresponding spectra were simultaneously recorded using a yttrium aluminum garnet (YAG) screen and a spectrometer. The power profiles were subsequently reconstructed utilizing the technique detailed in Ref.~\cite{behrens2014few, Zeng2022Online}. The measurements are presented in Fig.~\ref{fig7:exp1}.

When the dispersion strength is relatively small, the FEL pulse consistently retains a quasi-Gaussian shape. The first signs of pulse splitting appear at $R_{56}=\SI{0.15}{mm}$, corresponding to a pulse duration of \SI{110}{fs} and a pulse energy of \SI{33.8}{\micro\joule}. As $R_{56}$ continues to increase, two well-separated sub-bunches begin be emerge. This phenomenon arises because the electrons at the center of the laser pulse are over-bunched, whereas those near the head and tail are optimally bunched. When $R_{56}$ approaches approximately \SI{0.22}{mm}, the pulse splits into three sub-bunches, with a pulse duration of about \SI{178}{fs} and a pulse energy of around \SI{12.8}{\micro\joule}. The recurrence of the central sub-bunch is attributed to the Bessel function reaching its second local maximum as $R_{56}$ increases. The simultaneously measured pulse spectra, depicted in the right panel of Fig.~\ref{fig7:exp1}, exhibit a splitting phenomenon analogous to that observed in the FEL pulse profiles.

\begin{figure}[!htb]
  \centerline{
  \includegraphics[width=0.5\textwidth]{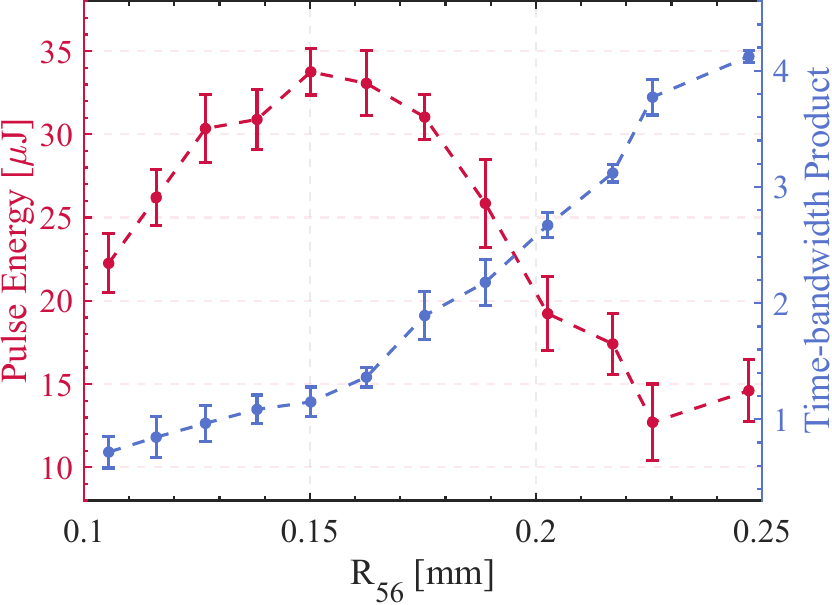}}
   \caption{Measured FEL pulse energy (red) and the time-bandwidth product (blue) at different dispersion strengths.}
   \label{fig8:exp2}
\end{figure}

Combining these two results, Figure~\ref{fig8:exp2} illustrates the evolution of the measured FEL pulse energy (red) and TBP (blue). As the dispersion strength increases, both the pulse duration and TBP experience consistent growth, while the pulse energy exhibits an upward and then downward trend. The initial TBP is approximately 0.72, which exceeds the Fourier limit by a factor of 1.6 due to the frequency chirp in the seed laser. The elongation of the seed laser pulses from $\sim$\SI{120}{fs} to $\sim$\SI{170}{fs} suggests the presence of this moderate frequency chirp, which is likely induced during the third harmonic generation and the laser transportation. Furthermore, a significant increase in the TBP is observed whenever the FEL pulse splits, whether transitioning from one into two or from two into three. 

The maximum of the FEL pulse energies takes place at the onset of pulse splitting, specifically at $R_{56}^{pe}\approx\SI{0.15}{mm}$. Since the maximum bunching factor occurs before pulse splitting, the corresponding dispersion strength, $R_{56}^{bun}$, will be necessarily smaller than $R_{56}^{pe}$. The optimal pulse energy stability is achieved at $R_{56}^{sta}\approx\SI{0.17}{mm}$, where the pulse duration is approximately \SI{134}{fs} and pulse energy is around \SI{31.2}{\micro\joule}. This indicates that $R_{56}^{bun}<R_{56}^{pe}<R_{56}^{sta}$, which aligns well with the results obtained from both theoretical analyses and numerical simulations. Additionally, the results depicted in Figure~\ref{fig8:exp2} indicate the potential necessity of balancing higher pulse energy with better temporal coherence, a consideration that is especially crucial in the context of cascaded seeded FELs.

Although experimental results can be affected by factors such as electron beam jitter and seed laser fluctuations, this experiment effectively demonstrates the influence of dispersion strength on the performance of seeded FELs. The measurements clearly illustrate the relationship among the required dispersion strengths that yield the maximum bunching factor, maximum pulse energy, and optimal pulse energy stability.

\section{Conclusion}\label{conclusion}

This study presents an alternative criterion for optimizing seeded FELs and provides a comprehensive investigation into the influence of dispersion strength on their performance through theoretical calculations, numerical simulations and experimental explorations. The effect of $R_{56}$ on various radiation characteristics in the HGHG scheme have been analyzed, including the power profile, spectrum, TBP, pulse energy, and stability. Additionally, we directly observed the phenomenon of pulse splitting and the corresponding spectral splitting by employing the approach outlined in Ref.~\cite{behrens2014few, Zeng2022Online}. A detailed analysis of the underlying mechanisms for these phenomena is also provided. Our findings indicate that a slight over-tuning of the dispersion strength can increase pulse energy and further improve pulse energy stability without significantly compromising temporal coherence in the HGHG scheme, adhering to the relationship $R_{56}^{bun}<R_{56}^{pe}<R_{56}^{sta}$. The results presented herein offer a more intuitive understanding of the seeded FEL lasing process and provide valuable insights for controlling the FEL pulse proprieties to improve the performance of the seeded FEL facilities.

~\\
\noindent\textbf{Declaration of Competing Interest}\\
The authors declared that they have no conflict of interest to this work.

~\\
\noindent\textbf{Acknowledgments}\\
The authors would like to thank Yifan Liang, Lingjun Tu, Hao Sun, Yong Yu, Duan Gu, Kaiqing Zhang, Bo Liu for useful comments on the simulations and experiments. This work is supported by the National Natural Science Foundation of China (Grant No. 12305359 and 22288201), the Scientific Instrument Developing Project of Chinese Academy of Sciences (Grant No. GJJSTD20220001).

~\\
\noindent\textbf{Data availability statement}\\
The data that support the results of this study are available from the authors upon reasonable request.





\end{document}